\documentclass[conference]{IEEEtran}
\IEEEoverridecommandlockouts
\usepackage{cite}
\usepackage{amsmath,amssymb,amsfonts}
\usepackage{algorithmic}
\usepackage{graphicx}
\usepackage{textcomp}
\usepackage{xcolor}
\usepackage{booktabs}
\usepackage{multicol}  
\usepackage{multirow} 
\usepackage{subfigure}
\def\BibTeX{{\rm B\kern-.05em{\sc i\kern-.025em b}\kern-.08em
    T\kern-.1667em\lower.7ex\hbox{E}\kern-.125emX}}
\begin{document}

\title{TDASS: Target Domain Adaptation Speech Synthesis Framework for Multi-speaker Low-Resource TTS\\
\thanks{$\ast$Corresponding author: Jianzong Wang (jzwang@188.com).}
}

\author{\IEEEauthorblockN{Xulong Zhang, Jianzong Wang$^{\ast}$, Ning Cheng, Jing Xiao}
\IEEEauthorblockA{\textit{Ping An Technology (Shenzhen) Co., Ltd., China} }
}
\maketitle

\begin{abstract}
% The current typical text-to-speech (TTS) techniques fail to perform well on differentiating each speaker's timbre information from the multi-speaker speech dataset. Especially, when the target speaker takes a small proportion of the dataset compared with other speakers, other speakers will severely influence the timbre and pitch of generated speech. This is a difficult problem in TTS. Inspired by animals' digestive process and metabolic activity after taking intake, we propose firstly a general bionic architecture, functional Digestive Metabolic Network (TDASS), which could exploit the traditional TTS model. The TDASS consists of a digestive network containing an encoder and attention mechanism, and a metabolic network including a self-interest classifier and
% supervised generator. It can ignore the dregs and then absorb the benefits of intake to generate the target compound. Simultaneously, we introduce the functional digestive enzyme concept to enhance the performance. Considering the above problem of TTS, we embed the Tacotron2 into the TDASS framework, as well as added the timbre feature as the digestive enzyme. The ablation experiment on the multi-speaker speech dataset shows that the proposed TDASS has improved greatly the performance of Tacotron2 on Mean Opinion Score (MOS) and Voice Similarity Score (VSS).
Recently, synthesizing personalized speech by text-to-speech (TTS) application is highly demanded. But the previous TTS models require a mass of target speaker speeches for training. It is a high-cost task, and hard to record lots of utterances from the target speaker. Data augmentation of the speeches is a solution but leads to the low-quality synthesis speech problem. Some multi-speaker TTS models are proposed to address the issue. But the quantity of utterances of each speaker imbalance leads to the voice similarity problem. We propose the Target Domain Adaptation Speech Synthesis Network (TDASS) to address these issues. Based on the backbone of the Tacotron2 model, which is the high-quality TTS model, TDASS introduces a self-interested classifier for reducing the non-target influence. Besides, a special gradient reversal layer with different operations for target and non-target is added to the classifier. We evaluate the model on a Chinese speech corpus, the experiments show the proposed method outperforms the baseline method in terms of voice quality and voice similarity.
\end{abstract}


\section{Introduction}

Text-to-speech synthesis (TTS) aims to generate intelligible and natural speech from the input text or phoneme sequence ~\cite{DBLP:conf/emnlp/MaZLZLP0020, DBLP:conf/icassp/ZengWCXX20,zhao2022nnspeech,DBLP:journals/corr/abs-1905-08459,aolan2021}. It has a long history in the TTS research, from the method of concatenative synthesis, and statistical parametric synthesis to the recent method of deep neural TTS. Concatenative synthesis methods and parametric synthesis methods are two main methods for TTS. Concatenative methods propose to combine each word's voice into a full-sentence voice. It means the method can make synthesis voice only if the database has all single words' representatives of the input text. The concatenative synthesis relies on the database of prestored utterance, it can synthesize natural speech nearly the same as the raw speech spoken by the person. But due to it needing a huge amount of utterances to cover different words, it is costly and cannot be used in a general way. Parametric synthesis methods transfer the known words and their voice to vectors and predict the unknown words' vectors. To use the known voice vectors and predicted vectors, parametric synthesis models can generate voice. While the statistical parametric synthesis method can address the drawbacks of the concatenative method by recovering the speech through limited acoustic parameters. But the quality of the statistical parametric synthesis has lower intelligible and robotic can be easily differentiated from human speech. 

Recent advances in TTS models achieved high-quality synthetic speech. Such as Tacotron series~\cite{DBLP:conf/interspeech/WangSSWWJYXCBLA17,DBLP:conf/icassp/ShenPWSJYCZWRSA18}, FastSpeech series~\cite{DBLP:conf/nips/RenRTQZZL19} are able to synthesis the approach human naturalness speech. But the challenges remain, these models require a number of clean speeches to train. But it is hard for the industry to collect the hundreds of speeches per person~\cite{DBLP:conf/icassp/HuybrechtsMCPSL21,zhang2022Singer,DBLP:conf/naacl/BansalKLLG19}. Furthermore, training one model for multiple speakers is another challenge (Multi-speaker TTS) to improve the efficiency of TTS models~\cite{DBLP:conf/icml/MinLYH21,qubo2021,DBLP:conf/slt/LiOLH21}.

Previous researches on multi-speaker TTS remain requires hundreds or thousands of high-quality training speeches per person~\cite{mitsui20_interspeech,zhang2021singer,chen20r_interspeech}. Several studies utilize voice conversion~\cite{zhou2021seen,zhang2020singing,kobayashi2021crank} to augment both speaker and speech databases to address the extensive training data required. But the recent advance of voice conversion is hard to synthesize noise-free speeches. These speeches restrict the outcome quality of downstream TTS tasks~\cite{DBLP:conf/icassp/Yan21,sibo2022,gao2021vocal,DBLP:conf/iclr/DonahueDBES21}. 

Most of the previous works utilize the timbre information of speakers for multi-speaker TTS. X-vector, which represents the speaker timbre features, is commonly used in multi-speaker TTS models~\cite{DBLP:conf/icassp/SnyderGSPK18,zhang2022MetaSID}. By adding the X-vector into Tacotron2~\cite{DBLP:conf/icassp/ShenPWSJYCZWRSA18}, it can synthesize speech for multiple speakers. However, the similarity of synthesis speech to the target speaker is another challenge. Because the synthesis process is easy to be influenced by the other speakers who are not the target. Some previous works believe the problem happens on the alignment, while others believe the network structure is unsuitable. The Light-TTS is a new SOTA model for multi-speaker TTS~\cite{DBLP:conf/icassp/LiOLH21,zhang2022MDCNN-SID}, which modifies the FastSpeech2~\cite{DBLP:conf/nips/RenRTQZZL19}. Cai \textit{et al.}~\cite{DBLP:conf/interspeech/CaiZL20} utilize a reconstruct timbre distance loss. Through control of the distance between the timbre embedding of synthesis speech to the original to restrict the voice similarity.

\begin{figure*}[!htb]
  \centering
  \includegraphics[width=0.9\linewidth]{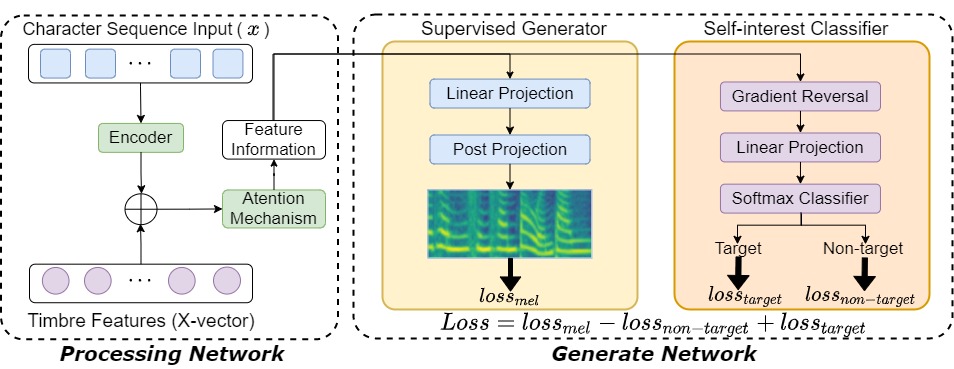}
  \caption{The architecture of the Target Domain Adaptation Speech Synthesis Network (TDASS).}
  \label{fig:flowchart}
\end{figure*}

We propose a new framework and training process for the multi-speaker TTS model to apply in the low-resource situation and address the voice similarity. The framework is called as \textbf{T}arget \textbf{D}omain \textbf{A}daptation \textbf{S}peech \textbf{S}ynthesis Network (TDASS), the structure is shown in Figure~\ref{fig:flowchart}. Based on the backbone of Tacotron2, we introduce a self-interested classifier to improve the synthesis of voice similarity. The self-interested classifier identifies and filters all the non-target encoding hidden vectors, which combine the linguistic and timbre information. Furthermore, a special gradient reversal layer (GRL) was added to the classifier. The GRL will give gradient reversal for non-target speakers and normal gradient for target speakers. Compared to the \cite{csmt2021sun,DBLP:conf/interspeech/CaiZL20}, rather than the reconstruct timbre distance loss, the classifier is more specific for the target speaker. We applied the TDASS to a database of Chinese speeches. The experiments show the proposed model has improved than the baseline method in terms of naturalness and similarity.

% need to be checked

\section{Related Work}
%  In this section, we revisited the related TTS works in traditional parametric synthesis methods and deep learning methods.

Hidden Markov Model (HMM) based TTS model is a statistic method of traditional parametric synthesis method \cite{DBLP:journals/speech/ZenTB09,zhang2022SUSing}. Using statistical results from the training dataset to reason and predict a possible voice used in the next state. HMM is a popular model used in synthesis voice \cite{DBLP:journals/pieee/TokudaNTZYO13,tang2022avqvc, DBLP:conf/icassp/TokudaYMKK00, asru2021tang,DBLP:journals/ijst/PatilL19, DBLP:journals/taslp/KoriyamaK19}. Because the HMM assumes the next state is only related to the current state. This is in line with the characteristics of human speech. However, the HMM-based synthesis methods predict accuracy is low to lead the generated voice worse than concatenative synthesis methods. Furthermore, the methods also required many databases to learn the relation between every single word. 

WaveNet is an autoregressive generative model, that predicts the next sampling point by learning the previous points \cite{DBLP:journals/corr/OordDZSVGKSK16}. WaveNet is similar to the HMM used in traditional parametric synthesis methods. It can not direct synthesize voice from the input text and require a preprocessing model to transfer text to features. The performance of TTS by using WaveNet is improved, but the result is influenced by the preprocessing. If the error occurs in the previous steps, the WaveNet is hard to fix it. Furthermore, WaveNet is slow to synthesize voice as it can only predict a sampling point per time.

With the development of deep model on many tasks~\cite{wang2022drvc,wang2020transfer,zhang2019novel}. Tacotron is the first end-to-end TTS model \cite{DBLP:conf/interspeech/WangSSWWJYXCBLA17}. Instead of using another model to preprocess the input text, the Tacotron can encode text and generate voice. The Tacotron consists of five parts, encoder, attention, decoder, and post-processing. It improves the performance in terms of WaveNet, which is the end-to-end model benefits. But the improvement is limited, as the output of Tacotron is mel-spectrogram and requires a model to transfer the mel-spectrogram to voice. Previous research uses the Griffin-Lim reconstruction algorithm, a simple model to generate voice but 
with little noise.

To address the problem of Tacotron, Tacotron2 is proposed, which is also an end-to-end model \cite{DBLP:conf/icassp/ShenPWSJYCZWRSA18}. Tacotron2 uses WaveNet vocoder to replace Griffin-Lim to synthesize voice from mel-spectrogram. Besides, Tacotron2 only generates one frame per step in the decoder and uses another post-net part to adjust the generated mel-spectrogram. 

Fastspeech~\cite{ren2019fastspeech} is not like the previous neural TTS, it firstly generated mel spectrum in a parallel way. FastSpeech is built up a feed-forward network based on Transformer, the attention alignments between encoder and decoder are achieved with phoneme duration prediction based on a teacher model. Through a length regulator to expand the source phoneme sequence to match the length of the target mel-spectrogram sequence for a parallel generation. Due to mel spectrum was relied on the distilling of the teacher-student model, there exists information loss. The Fastspeech2~\cite{ren2020fastspeech} was proposed to do a enhance of the Fastspeech, it introduces variance adaptor to enhance the variation information such as pitch, energy, and duration. The training avoids two-stage teacher-student distillation, it directly uses the groundtruth mel spectrum.

While the previous neural TTS modes need two-stage, one is the acoustic model to generate mel spectrum from linguistic information, and the other is the vocoder to generate audio wav from mel spectrum. Several recent neural TTS models enabling single-stage training and parallel sampling have been proposed. VITS~\cite{kim2021conditional} adopts normalizing flows and an adversarial training process for variational inference augmentation, which improves the expressive of the generator. A stochastic duration predictor to synthesize speech with diverse rhythms from input text was added in VITS, it could encode the same input text in multiple ways for speaking.

But training the high-quality TTS systems needs a large amount of paired text and speech data. Most languages lack training data for developing TTS systems. The low resources TTS works often utilize self-supervised training, cross-lingual transfer, cross speaker transfer, and dataset augmentation in the wild.

% need to be checked

\section{Proposed Method}

% In this paper, the TDASS is a speech synthesis framework aimed to address the synthesis voice similarity problem. The overall framework of our TDASS is shown in Figure~\ref{fig:flowchart}. The TDASS for TTS is described in detail in this section.

% The proposed network is a new bionic concept, which can be applied to different tasks such as data expansion and filtering. The network is composed of a digestive network and a metabolic network in Figure~\ref{fig:flowchart}.
\subsection{Backbone Modules}
The processing network consists of an encoder and attention mechanism. The primary purpose of the processing network is to encode the input embedding of the phoneme sequence and decode the high-level feature added with speakers' timbre feature to each frame of mel-spectrogram.

Given the phoneme sequences $x$, the output $z$ of encoder is calculated as $z = \phi (x,\theta_p)$, where $\phi (\cdot)$ is the mapping  function of encoder. $\theta_p$ is the parameter of encoder. Then, we concatenate the phoneme embedding $z$ and timbre $x\text{-}vector$ to get the embedded features, $\mathcal{Z}$.
% \begin{equation}
%     z = \phi (x,\theta_p)
% \end{equation}
% \noindent where $\phi (\cdot)$ is the mapping  function of encoder. $\theta_p$ is the parameter of encoder.
% To embedded the timbre feature $x\text{-}vector$, which is extracted from the corresponding speech sample,  into $z$ as follows:
% \begin{equation}
%     \mathcal{Z} = concat(z, x\text{-}vector)
% \end{equation}
% \noindent Where $\mathcal{Z}$ is the input of attention mechanism.

The attention mechanism takes the embedded features, $\mathcal{Z}$, as input. The output features of the processing network, $\delta$, is calculated as $\delta  = f(\mathcal{Z},\theta_d)$, where $f (\cdot)$ is the mapping function of the attention mechanism. $\theta_d$ is the parameter of the attention mechanism. 
% The processing network decomposes the original context-sensitive feature in $\delta$ into each element of the mel-spectrogram.
% \begin{equation}
%     \delta  = f(\mathcal{Z},\theta_d)
% \end{equation}
% \noindent Where $f (\cdot)$ is the mapping function of the attention mechanism. $\theta_d$ is the parameter of the attention mechanism. The processing network decomposes the original context-sensitive feature in $\delta$ into each element of the mel-spectrogram.

% The Generate network is composed of a self-interest classifier and supervised generator. After receiving the feature information, $\delta$, of the processing network, the generate network aims to generate the voice. To begin with, it aims to remove the timbre information of the non-target speaker by using the self-interest classifier. Then, the supervised generator can generate the mel-spectrogram under the guidance of the target speaker's timbre information. The supervised generator is designed the same as Tacotron2. We will thoroughly discuss the self-interest classifier in this paper, which is a new part of Tacotron2.
The Generative network is composed of a self-interest classifier and supervised generator. The generator is the same as the decoder of Tacotron2~\cite{DBLP:conf/icassp/ShenPWSJYCZWRSA18}. It aims to synthesize the mel-spectrogram for the input feature information, $\delta$. 

\subsection{Self-interested Classifier}
We introduce the self-interested classifier $\mathcal{C}$ takes the feature information $\delta$ and passes it through three fully connected layers and a Softmax function. The Softmax function identifies whether the voice is spoken by the target or non-target. The output is two probabilities $\mathcal{P}_0,\mathcal{P}_1$ for non-target and target speakers.
\begin{equation}
(\mathcal{P}_0,\mathcal{P}_1)=\mathcal{C}\left(\delta ; \theta_{C}\right)=\mathcal{C}\left(\mathcal{P}\left(x, x-vector ; \theta_{P}\right) ; \theta_{C}\right)
\end{equation}
\noindent where $\mathcal{C}$ denotes the self-interest classifier, and the parameters of the self-classifier is labeled as $\theta_C$. The parameter of the processing network is $\theta_P$, which equal with $(\theta_p, \theta_d)$. The self-interest classifier $\mathcal{C}$ and the generate network are trained together with the cross entropy loss on speaker classification:
\begin{equation}
    \begin{split}
        \mathcal{L}_{\text{CLS}}(\theta_{P}, \theta_{C})&=\mathcal{L}_{non-target}+\mathcal{L}_{target},\\
        \mathcal{L}_{non-target} &= -\Pi(y_{speaker}==0)\log \mathcal{P}_0,\\
        \mathcal{L}_{target} &= -\Pi(y_{speaker}==1)\log \mathcal{P}_1.
    \end{split}
\end{equation}
% \begin{equation}
%     \begin{split}
%         \mathcal{L}_{\text{CLS}}(\theta_{P}, \theta_{C})&=loss_{non-target}+\frac{1}{\lambda+loss_{target}},\\
%         loss_{dregs} &= -\Pi(y_{speaker}==0)\log P_0,\\
%         loss_{target} &= -\Pi(y_{speaker}==1)\log P_1.
%     \end{split}
% \end{equation}

% \begin{equation}
%     \begin{split}
%         \begin{array}{c}
%         \mathcal{L}_{\text{CLS}}(\theta_{e}, \theta_{c})=\sum_{k=1}^{k=k^{\prime}-1}\Pi(y_{\text {speaker}}=k) \log P_{k}\\
%          &+\sum_{\mathrm{k}=k^{\prime}-1}^{k=K} \Pi(y_{\text {speaker}}=k) \log P_{k}\\
%          &-\Pi(y_{\text{speaker}}=k^{\prime}) \log P_{k^{\prime}}
%         \end{array}
%     \end{split}
% \end{equation}

\noindent where the indicator function is labeled as $\Pi(\cdot)$, speech $x$ is spoken by the speaker of $y_{speaker}$, for non-target speakers the classification loss is labeled as $\mathcal{L}_{non-target}$, and $\mathcal{L}_{target}$ denotes the target speaker's classification loss. 

During training, parameters $\theta_C$ are used to discriminate the target and non-target speaker by minimizing the classification loss, whereas gradient reversal is used to update $\theta_P$. During the minimization of classification and maximization of network generation, it will be converged when the output is similar to the target speaker. Finally, the non-target speaker could not be identified by the classifier.

The supervised generator $G$ and the processing network are jointly with the self-interest classifier $\mathcal{C}$ for training a multi-task learning framework,
\begin{align}
\begin{split}
\mathcal{L}(\theta_P, \theta_G, \theta_C)&= \mathcal{L}_{GLS}(\theta_P, \theta_G) \\
&- \lambda \mathcal{L}_{non-target}(\theta_P, \theta_C) \\
    &+\mathcal{L}_{target}(\theta_P, \theta_C)
\end{split}
\label{eq5}
\end{align}
% \begin{equation}
% \mathcal{L}(\theta_P, \theta_G, \theta_C)=\mathcal{L}_{GLS}(\theta_P, \theta_G)+F(\mathcal{L}_{CLS}(\theta_P, \theta_C))
% \label{eq5}
% \end{equation}
\noindent where $\theta_G$ is the parameters of the supervised generator, $\mathcal{L}_{GLS}(\theta_P, \theta_G)$ is the distance between generated mel spectrogram and ground truth. Parameters $\theta_P$, $\theta_C$, $\theta_G$ are 
% \begin{gather}
%     \theta_e, \theta_P=\arg\min\mathcal{L}(\theta_e, \theta_P, \theta_c)\\
%     \theta_c=\arg\min\mathcal{L}(\theta_e, \theta_P, \theta_c)
% \end{gather}
% \begin{equation}
%     \theta_P, \theta_G=\arg\min\mathcal{L}(\theta_L, \theta_P, \theta_C)
% \end{equation}
% \begin{equation}
%     \theta_C=\arg\max\mathcal{L}(\theta_L, \theta_P, \theta_C)
% \end{equation}
updated though back-propagation.
% \begin{equation}
% \theta_{P} \leftarrow \theta_{P}-\mu\left(\frac{\partial \mathcal{L}_{GLS}}{\partial \theta_{P}} + F(\frac{\partial \mathcal{L}_{CL S}}{\partial \theta_{P}})\right)
% \end{equation}
% \begin{equation}
% \theta_{G} \leftarrow \theta_{G}-\mu \frac{\partial \mathcal{L}_{GLS}}{\partial \theta_{G}}
% \end{equation}
% \begin{equation}
% \theta_{C} \leftarrow \theta_{C}-\mu \frac{\partial \mathcal{L}_{CLS}}{\partial \theta_{C}}
% \end{equation}
% \noindent where $\mu$ is the learning rate. Due to the gradient reversal layer, the gradient reversal maximizes $\mathcal{L}_{non-target}$ in $\mathcal{L}_{CLS}$ for $\theta_P$ and minimizes $\mathcal{L}_{target}$ in $\mathcal{L}_{CLS}$ while $\mathcal{L}_{GLS}$ is always minimized for optimize the $\theta_P$ and $\theta_G$.

\subsection{Target Domain Adaptation: Special Gradient Reversal Layer}

For synthesis voice similarity and target domain adaptation, in self-interest classifier. For target and non-target, we process a different way of gradient reversal \cite{ganin2015unsupervised,zhang2020research}. The self-interest classifier has a specific designed gradient back propagated method as follow:
\begin{equation}
    \begin{split}
        F(\frac{\partial \mathcal{L}_{CL S}}{\partial \theta_{P}}) =  -\lambda \frac{\partial \mathcal{L}_{non-target}}{\partial \theta_{P}} +  \frac{\partial \mathcal{L}_{target}}{\partial \theta_{P}}
    \end{split}
    \label{gradient}
\end{equation}

% \begin{equation}
%     \begin{split}
%         F(\mathcal{L}_{\text{CLS}}) &=  F(\mathcal{L}_{non-target}) +  F(\mathcal{L}_{target}),\\
%         F(\mathcal{L}_{non-target}) &= -\mathcal{L}_{non-target},\\
%         F(\mathcal{L}_{target}) &= -\frac{1}{\lambda+loss_{target}}
%     \end{split}
%     \label{gradient}
% \end{equation}
\noindent where $F(\cdot)$ is the mapping function of gradient reversal layer. $\lambda$ is the weight adjustment parameters.
% For non-target voice, we set negative $\lambda$ for back-propagation. For the target voice, we set the $\lambda$ as -1. 
The processing network parameters $\theta_{P}$ are updated based on the gradient back-propagation loss function. 
% When the training sample is non-target, the back-propagation of $\theta_P$ is calculated as follow:
\begin{equation}
\small
\theta_{P} \leftarrow \theta_{P}-\mu\left(\frac{\partial \mathcal{L}_{GLS}}{\partial \theta_{P}}  -\lambda \frac{\partial \mathcal{L}_{non-target}}{\partial \theta_{P}} + \frac{\partial \mathcal{L}_{target}}{\partial \theta_{P}}\right)
\label{D_S}
\end{equation}
When the training sample is target, the back-propagation of $\theta_P$ is calculated as follow:
\begin{equation}
\theta_{P} \leftarrow \theta_{P}-\mu\left(\frac{\partial \mathcal{L}_{GLS}}{\partial \theta_{P}}  + \frac{\partial \mathcal{L}_{target}}{\partial \theta_{P}}\right).
\label{D_G}
\end{equation}

When the training sample belongs to a non-target, the $\lambda \frac{\partial \mathcal{L}_{non-target}}{\partial \theta_{P}}$ equal to zero. It aims to gradually removed the style influence from the non-target speakers under adversarial training. The $\frac{\partial \mathcal{L}_{target}}{\partial \theta_{P}}$ is zero, when the training sample comes from the target speaker. The operation optimize towards the style information of the target speaker under supervised training.

In the TDASS, we set the loss of the self-interest classifier as the style loss, representing that the timbre information of embedded feature information is target or non-target. Besides, the loss of supervised generator is primarily composed of the style loss and the content loss, which aim the pronunciation of input text similar to the real. Therefore, the style information of the non-target speakers,  in Equation (\ref{D_S}), is gradually removed under adversarial training. In Equation (\ref{D_G}), the feature information, $\theta$, is optimized towards the style information of the target speaker under supervised training.  The proposed training process makes the processing network remove the non-target information, and the generated network generates the mel-spectrogram for target speakers. 

\label{Section2}

\begin{table*}[htp]
  \centering
  \fontsize{8}{7}\selectfont
  \caption{The MOS and VSS of different models in ablation study. $\Downarrow$ means lower score is better, and $\Uparrow$ means bigger score is better. The number of utterances means how many the target speaker speech is included in the fine-tune process.}
  \label{MOS_VSS}
    \begin{tabular}{ccccccccc}
    \toprule
    \multirow{2}{*}{Method}&
    \multicolumn{2}{c}{ 30 utterances}&\multicolumn{2}{c}{ 100 utterances}&\multicolumn{2}{c}{ 300 utterances}&\multicolumn{2}{c}{ 500 utterances}\cr
    \cmidrule(lr){2-3} \cmidrule(lr){4-5} \cmidrule(lr){6-7} \cmidrule(lr){8-9}
    & MOS $\Uparrow$ & MCD (dB) $\Downarrow$ & MOS $\Uparrow$ & MCD (dB) $\Downarrow$& MOS $\Uparrow$ & MCD (dB) $\Downarrow$& MOS $\Uparrow$ & MCD (dB) $\Downarrow$\cr
    \midrule
    Ground truth & 4.53 $\pm$ 0.05
    & --- 
    & --- & ---
    & --- & --- 
    & --- & --- \cr
    Tacotron2 $w/o$ X-vector & 2.83 $\pm$ 0.24 & 9.49 $\pm$ 0.35 
    & 3.08 $\pm$ 0.18 & 9.72 $\pm$ 0.33 
    & 3.42 $\pm$ 0.14 & 9.81 $\pm$ 0.64
    & 3.70 $\pm$ 0.13 & 9.99 $\pm$ 0.46 \cr
    Tacotron2 & 3.06 $\pm$ 0.22 & 9.60 $\pm$ 0.30
    & 3.29 $\pm$ 0.16 & 9.52 $\pm$ 0.35
    & 3.53 $\pm$ 0.15 & 9.71 $\pm$ 0.27
    & 3.84 $\pm$ 0.13 & 9.99 $\pm$ 0.56\cr
    \midrule
    TDASS $w/o$ X-vector & \underline{3.38 $\pm$ 0.18} & \underline{9.39 $\pm$ 0.33}
    & \underline{3.58 $\pm$ 0.15} & \underline{9.47 $\pm$ 0.32}
    & \underline{3.76 $\pm$ 0.15} & \textbf{ 9.40 $\pm$ 0.46}
    & \underline{3.90 $\pm$ 0.11} & \textbf{9.36 $\pm$ 0.23}\cr
    TDASS & \textbf{3.51 $\pm$ 0.16} & \textbf{9.36 $\pm$ 0.30}
    & \textbf{3.70 $\pm$ 0.13} & \textbf{9.47 $\pm$ 0.14}
    & \textbf{3.86 $\pm$ 0.10} & \underline{9.45 $\pm$ 0.65}
    & \textbf{3.98 $\pm$ 0.08} & \underline{9.60 $\pm$ 0.26}\cr
    \bottomrule
    \end{tabular}
\end{table*}

\section{Experiments}

To validate the performance of our proposed method, we design an ablation study on a small-capacity target speaker dataset of 30, 100, 300, 500. In the experiment, we compare the TDASS with the baseline model of Tacotron2, and the baseline model with adding timbre feature of x-vector and TDASS without timbre feature in detail.

\subsection{Datasets}

We evaluate the proposed framework on a dataset of Chinese speech corpus. There are three speakers with 4019, 20257, 8915 records, respectively, in the database. The sampling rate of the audios is 22050 Hz. We set speaker1 and speaker2 are non-target speakers and used them for training the TDASS without the self-interested classifier. The total 24276 speeches of the two speakers are split into 24000 training records and 276 validation records. Then, we utilize speaker3's speeches and the previous speakers to fine-tune the TDASS with the self-interested classifier. To show the low-resource performance of the proposed model, we set four experiments that randomly select 30, 100, 300, 500 utterances from speaker3, respectively. The test database of the four experiments uses the same 100 samples.

% We carry out the experiments on a private commercial dataset collected in Chinese, and the sampling rate of the audio data is 22050. The total number of Chinese audio clips and the corresponding text transcripts is 33191, with which the total length is 30 hours approximately. This dataset contains three speakers, speaker1, speaker2, speaker3, of which there are 4019, 20257, 8915 records, respectively. In this experiment, we set speaker3 as the target.

% For pre-training, we aim to train the Tacotron2 as we build the TDASS based on it. Therefore, we randomly split the speaker1 and speaker2 voices into 24000 training records and 276 validation records. 

% For the process of synthesis voice, we randomly respectively get 30, 100, 300, 500 sample records from speaker3 as training data. The data will mix the speaker 1 and speaker 2 voices to input the TDASS. Furthermore, the corresponding validation set has 100 records.

\begin{figure*}[!htb]
\centering
\subfigure[Ground Truth]{\label{fig:subfig:a}
\includegraphics[width=0.18\linewidth]{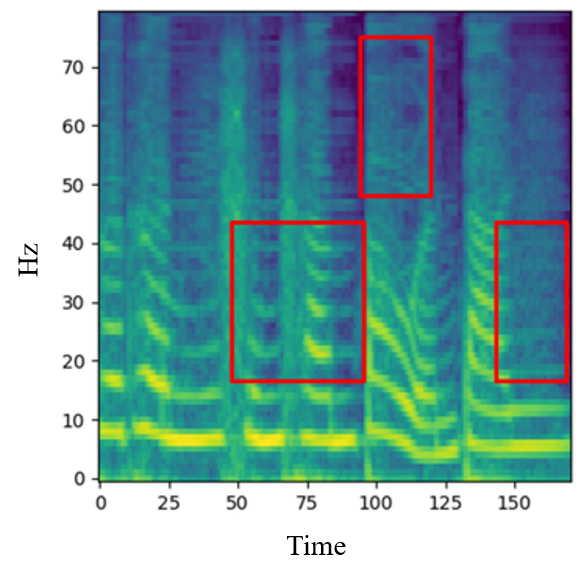}}
\hspace{0.05\linewidth}
\subfigure[Tacotron2]{\label{fig:subfig:b}
\includegraphics[width=0.18\linewidth]{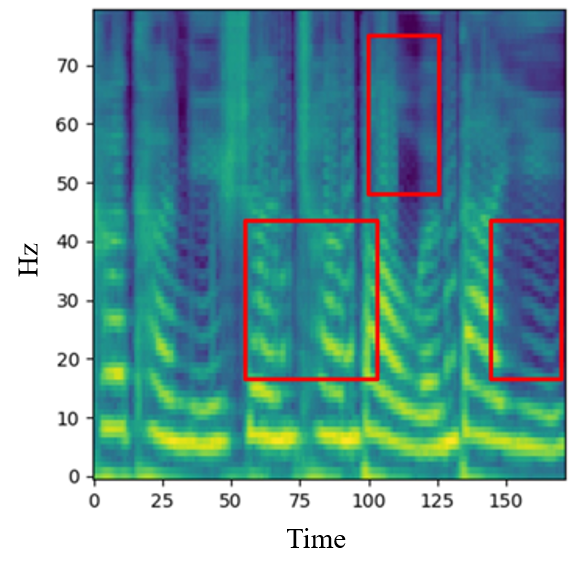}}
% \vfill
\hspace{0.05\linewidth}
\subfigure[TDASS $w/o$ X-vector]{\label{fig:subfig:c}
\includegraphics[width=0.18\linewidth]{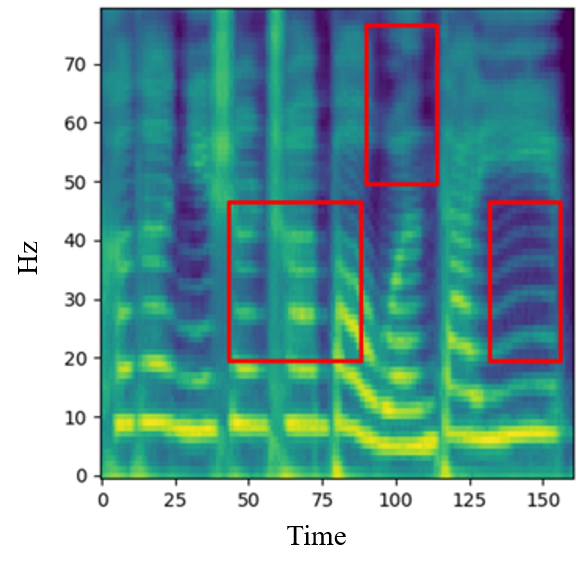}}
\hspace{0.05\linewidth}
\subfigure[TDASS]{\label{fig:subfig:d}
\includegraphics[width=0.18\linewidth]{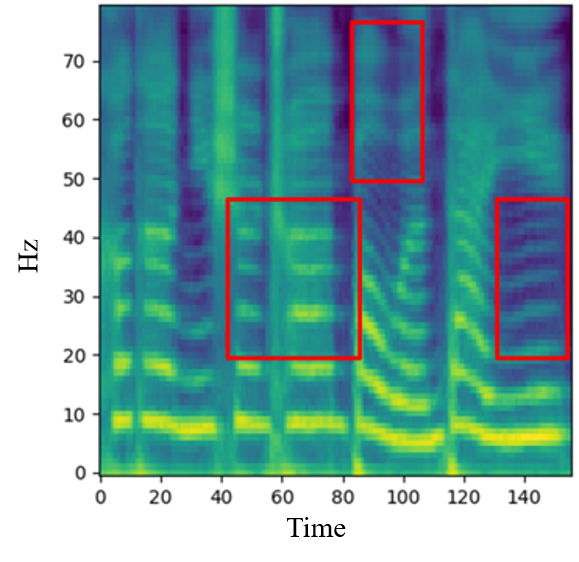}}
\caption{The comparison on similarity between ground truth, Tacotron2, TDASS without X-vector and TDASS. The red box region shows the significant difference between them.}
\label{fig:melspec}
\end{figure*}
\subsection{Configuration}

The backbone configuration are following the Tacotron2. To specifically specify is that, the self-interest classifier contains three fully connected layers, of which the set are (1536,1024), (1024,64), and (64,2), respectively. The timbre feature $x\text{-}vector$ is a one-dimensional vector obtained from each training sample and the length of the vector is 512.

Our proposed model was trained on a single NVIDIA V100 GPU. To begin with, we pre-train the TDASS without a self-interested classifier on the non-target speakers, speaker1 and speaker2, utterances. Then we load the parameter into the TDASS with a self-interested classifier to synthesize the target voice on the multi-speakers, all three speakers. Following \cite{ganin2015unsupervised}, we gradually changed the parameter $\lambda$ in the self-interest classifier from 0 to 1 as follow
\begin{equation}
    \lambda = \frac{2}{1 + exp(-10 \cdot k)} - 1
\end{equation}
\noindent where, $k$ is the percentage of the training process. We train our model with batch size of 24 samples, and use the default Adam optimizer with $\beta_1=0.9, \beta_2=0.999, \varepsilon=10^{-8}$. We adopt the learning rate of $10^{-4}$ and apply $L_2$ regularization with weight $10^{-6}$\cite{DBLP:conf/icassp/ShenPWSJYCZWRSA18,asru2021zhang}.

\subsection{Evaluation Metrics}

We utilize subjective and objective well-known metrics to evaluate the Tacotron2 $w/o$ X-vector, Tacotron2, TDASS $w/o$ X-vector, and TDASS.%\footnote[1]{\href{https://tts-sci-zhangxulong.github.io/FDMN-TTS/}{Available at https://tts-sci-zhangxulong.github.io/FDMN-TTS/}.}. 

\textbf{Objective metrics} Mel cepstral distortion (MCD) is one of the objective metrics widely used to compare two speeches' similarities. It represents the distance between the Mel-scale Frequency Cepstral Coefficients (MFCC) of synthesis speech and the original speech's MFCC. Besides, we also provide an example of mel-spectrograms comparison, shown in Fig. \ref{fig:melspec}. The more similar mel-spectrogram, the more similar voice. Because the mel-spectrogram is obtained by passing the spectrogram, which is transferred from voice by short-time Fourier transformation (STFT),  through the mel-scale filter banks. It means the mel-spectrogram can be seen as the voice shown in two-dimensional visualization.

% Besides, we also compare the mel-spectrograms of ground truth, the previous model, and the proposed model. The purpose is to visualize similarity comparison. The more similar mel-spectrogram, the more similar voice. Because the mel-spectrogram is obtained by passing the spectrogram, which is transferred from voice by short-time Fourier transformation (STFT),  through the mel-scale filter banks. It means the mel-spectrogram can be seen as the voice shown in two-dimensional visualization.

\textbf{Subjective metrics} Mean Opinion Score (MOS) and Voice Similarity Score (VSS)~\cite{DBLP:journals/corr/abs-2004-11490,DBLP:journals/csl/ViswanathanV05} are used in this experiment. MOS means to identify whether the converted voice is clear or not. VSS aims to determine the most similar to the real voice.

Both MOS and VSS are obtained by inviting native speakers to rate for the synthesis audio. We have 30 testers with an equal number of men and women. Respondents in the test have various knowledge backgrounds, such as Nature Language Processing, Product Manager, Psychology, etc. 

For MOS evaluation, we give the testers a total of 140 utterances, 40 speeches per experiment, and eight samples per model of each experiment. Testers score zero to five for given samples, and higher marks mean close to naturalness. For VSS evaluation, we categorize the speeches with the same content into a group. Each group has one ground truth voice and four synthesis samples. Testers need to give zero to five marks for the synthesis speech too. %The higher marks mean more similar to ground truth.

\begin{figure}
    \centering
    \includegraphics[width=1\linewidth]{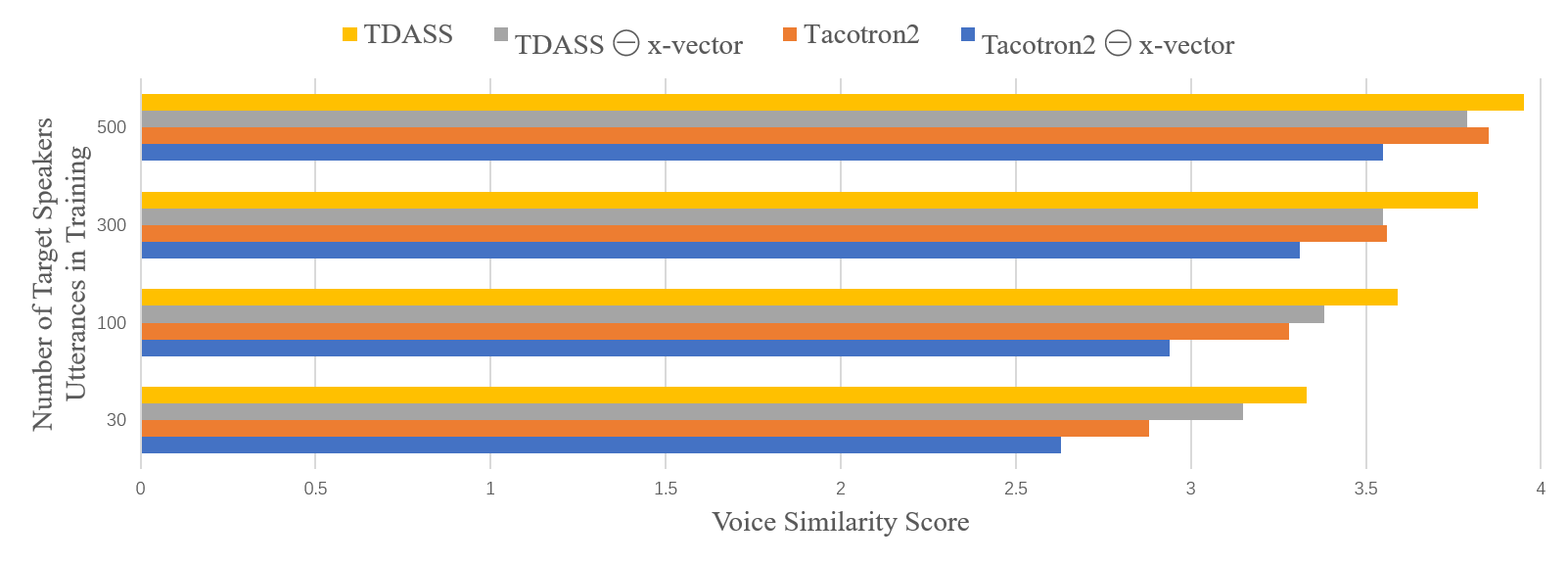}
    \caption{The voice similarity of different models in different training process.}
    \label{fig:vss}
\end{figure}

\subsection{Result and Discussion}
The MOS and MCD results for each model are shown in Table~\ref{MOS_VSS}. The VSS result is shown in Figure~\ref{fig:vss}.

\textbf{Overall} The proposed TDASS has better performance whether the low-resource situation and in terms of the different evaluation metrics. The TDASS shows the improvement in the low-resource condition. Furthermore, Fig. \ref{fig:melspec} shows the proposed model has most similar to the mel-spectrogram of the ground truth speech.

\textbf{Low-resource analysis} The training sample number of target speaker utterances was tested using 30, 100, 300, and 500 in experiments to fine-tune the pretrained model for the low-resource tests. The TDASS improves about 0.2 on MCD, 0.7 on MOS, and 0.6 on VSS on comparison of baseline method in these tests. The higher marks prove the TDASS improves the synthesis voice similarity and the naturalness and is listenable in low-resource. Besides, when the number of target speeches in fine-tune increases, all four models improve the synthesis speech performance. Table~\ref{MOS_VSS} and Figure~\ref{fig:vss} show the Tacotron2 improved about 0.8 points on both MOS and VSS from 30 utterances used to 500 samples, which satisfies the instinct. When there are 500 target speeches in fine-tune, the Tacotron2 approaches the normal performance with the 3.8 MOS scores. But in the meantime, the proposed model has better performance than Tacotron2. It proves whether the speaker utterances are limited, the proposed model has outperformance.

But one of the interesting findings is that when the number of target speaker utterances increases, the MCD value is higher. Besides, the TDASS $w/o$ X-vector has better performance on the MCD in the 300 and 500 utterances tests. Nevertheless, MOS and VSS still show the synthesis speech of TDASS is more clear and similar. Because the MCD is an objective metric for comparing the MFCC. The differences of MFCC may not lead the synthesis speeches unclear and unlike voice of the target speaker.

\textbf{Self-interested classifier analysis} The difference between Tacotron2 $w/o$ X-vector and TDASS $w/o$ X-vector is the self-interested classifier. But the experiments show there is an average of 0.5 points on the MOS and 0.6 points on the VSS improvements. It means even without the speaker timbre features, the self-interested classifier also improves the quality of Tacotron2 synthesized speeches. TDASS $w/o$ X-vector 0.3 points higher on MOS in the low-resource situation than Tacotron2 and 0.1 points larger in the normal condition. The comparison shows the different abilities based on the self-interested classifier and x-vector to enhance the synthesis speech quality. Because the Tacotron2 utilizes x-vector, TDASS $w/o$ X-vector did not. But TDASS $w/o$ X-vector uses the proposed self-interested classifier. Two comparisons prove the self-interested classifier has significant help for the multi-speaker TTS task.

 \textbf{Ablation Experiment about X-vector}. The training process is the same. Table~\ref{MOS_VSS} shows whether X-vector uses or not, the TDASS has certain improvements for traditional Tacotron2 under different conditions. But both the TDASS with $x\text{-}vector$ and Tacotron2 with $x\text{-}vector$ enhance the performance for each condition than without it. It means the X-vector is useful to improve the performance of models.

% \subsubsection{Comparison of Mel-spectrogram on Similarity}
% % We design the gradient reversal layer to help the digestive network learn the target feature while adding the gradient reversal after the attention mechanism.
% From the pixels shown in the three red box, we can find that the frame's distribution from our model is more similar to the ground truth, which demonstrates effectiveness in guiding the digestive network to generate the speech with target timbre.

In a word, the proposed framework, TDASS, and self-interested classifier are positive for synthesizing high-quality and high-similarity speeches. The $x\text{-}vector$, also has little significance on the generated speech.

% when the target speaker takes a small proportion of the dataset, TDASS framework improves significantly the performance of the Tacotron2, as well as the digestive enzyme $x\text{-}vector$, also has a little significance on the generated speech. We show the representation of the results on GitHub.

\section{Conclusions}
To address the low-resource and synthesis speech similarity of the multi-speaker TTS task, we propose the TDASS framework. The TDASS mainly contains a backbone of an acoustic module based on Tacotron2, a self-interested classifier, and a timbre embedding of X-vetor. It could reduce the non-target speaker influence and adapt to the target speaker domain, which benefits from the proposed self-interested classifier. The X-vector is used for the speaker embedding during the encoder to a latent variable, and TDASS first is trained based on the backbone of the Tacotron2-based module, then fine-tune with the proposed self-interested classifier. We set four experiments with different quantities of target utterances, the low-resource parts utilize 30 and 100 samples, and the normal parts have 300 and 500 speeches. The experiment results on a Chinese speech corpus show the proposed method outperforms the baseline methods even under the low-resource condition.

\section{Acknowledgement}
This paper is supported by the Key Research and Development Program of Guangdong Province under grant No.2021B0101400003. Corresponding author is Jianzong Wang from Ping An Technology (Shenzhen) Co., Ltd (jzwang@188.com).
% \section{Acknowledgements}

% This paper is supported by National Key Research and Development Program of China under grant No. 2017YFB1401202,
% No. 2018YFB1003500, and No. 2018YFB0204400.
\bibliographystyle{IEEEtran}
\bibliography{mybib.bib}

\end{document}